\documentclass[aps,prd,amssymb,amsmath,showpacs,preprintnumbers,nofootinbib]{revtex4-1}

\usepackage{graphicx}
\usepackage{epsfig}
\newcommand{\ba}{\begin{eqnarray}}
\newcommand{\ea}{\end{eqnarray}}

\newcommand{\be}{\begin{equation}}
\newcommand{\ee}{\end{equation}}
\newcommand{\p}{\partial}

\newcommand{\al}{\alpha}
\newcommand{\ga}{\gamma}

\newcommand{\Ga}{\Gamma}

\newcommand{\da}{\delta}
\newcommand{\Da}{\Delta}

\newcommand{\za}{\zeta}
\newcommand{\sa}{\sigma}
\newcommand{\en}{\varepsilon}
\newcommand{\oa}{\omega}
\newcommand{\Oa}{\Omega}
\newcommand{\la}{\lambda}

\newcommand{\n}{\nabla}

\newcommand{\ad}{\dot{\alpha}}
\newcommand{\bd}{\dot{\beta}}

\newcommand{\au}{i}
\newcommand{\bu}{j}

\newcommand{\+}{^{\dagger}}
\renewcommand{\vec}{\mathbf}

\newcommand{\w}{\widetilde}

\newcommand{\pv}{\vec{p}}
\newcommand{\se}{\bar{\sa}}
\newcommand{\cD}{{\cal D}}

\newcommand{\cO}{{\cal O}}

\newcommand{\cL}{{\cal L}}
\newcommand{\cM}{{\cal M}}

\newcommand{\2}{\frac{1}{2}}

\newcommand{\4}{\frac{1}{4}}

\newcommand{\stwo}{\sqrt{2}}

\newcommand{\ra}{\rightarrow}
\newcommand{\Ra}{\Rightarrow}

\newcommand{\LF}{\left(}
\newcommand{\RF}{\right)}
\newcommand{\LT}{\left[}
\newcommand{\RT}{\right]}

\newcommand{\Eq}[1]{(\ref{#1})}

\def\rme{e}
\def\rmd{d}
\def\rmi{i}

\begin{document}
\title{Cosmological Bardeen--Cooper--Schrieffer condensate as dark energy}

\author{Stephon Alexander}
\affiliation{Department of Physics and Astronomy,
Haverford College, Haverford, Pennsylvania 19041, USA}

\author{Tirthabir Biswas}
\author{Gianluca Calcagni}
\affiliation{Institute for Gravitation and the Cosmos, Department of Physics,\\ The Pennsylvania State University, 104 Davey Lab, University Park, Pennsylvania 16802, USA}

\date{June 28, 2009}

\begin{abstract}
We argue that the occurrence of late-time acceleration can conveniently be described by first-order general relativity covariantly coupled to fermions. Dark energy arises as a Bardeen--Cooper--Schrieffer condensate of fermions which forms in the early universe. At late times, the gap and chemical potential evolve to have an equation of state with effective negative pressure, thus naturally leading to acceleration.
\end{abstract}

\preprint{arXiv:0906.5161 \hspace{2cm} IGC-09/6-3}
\preprint{PHYSICAL REVIEW D {\bf 81}, 043511 (2010); {\bf 81}, 069902(E) (2010)}\hfill

\maketitle


\section{Introduction}

Current cosmological observations point to a universe dominated by a negative-pressure fluid component, dubbed dark energy, whose origin is unknown. If the equation of state of this fluid does not evolve and is $w =p/\rho=-1$, then dark energy is a cosmological constant. In this case, observations give a value of the cosmological constant that is 120 orders of magnitude smaller than the theoretically expected evaluation. We still lack a convincing way of understanding this issue. As a result, an approach towards making progress is to assume that the cosmological constant/dark energy stem from some new physics.\footnote{For alternative approaches which try to avoid dark energy by invoking large scale inhomogeneities see, for instance, \cite{void1,void2,reviews}.}

Ever since the observational evidence of late-time acceleration, model builders have sought to find a candidate for dark energy. However, this task is daunting as it is hard to identify an existing degree of freedom in the standard model or general relativity that (i) has negative pressure, (ii) is homogeneous on horizon scales and (iii) matches the observed energy scale. This initially led to quintessence models where a new fundamental scalar degree of freedom with a fine-tuned potential can be adjusted to yield a late-time acceleration tracking dark matter. Alternatively, investigators have invoked infrared modifications of general relativity, for instance $f(R)$ or Gauss--Bonnet gravity. While successful, many representatives in both classes of models suffer from fine-tuning or other theoretical problems due to the introduction of new degrees of freedom.

In this work, we take a minimalistic approach to dark energy assuming no extra degrees of freedom except fermionic matter on a flat Friedmann--Lema\^itre--Robertson--Walker (FLRW) background. A finite density of fermions in the early universe can undergo a Bardeen--Cooper--Schrieffer (BCS) condensation due to a covariant attractive channel from general relativity. The system is described by a set of transcendental equations that give relations between the scale factor, the fermion gap and the chemical potential. It was already shown in \cite{BS,Sha01,AB1} that such a BCS condensate can play an important role in the early universe by resolving the big bang singularity via a bounce. Remarkably, we find that depending upon some of the parameters of the theory, the same condensate can also affect the history of the universe at late times. In this paper we will show that the nonperturbative potential of the fermion gap can lead to late-time acceleration.\footnote{On the other hand, the present proposal does not address either the smallness or coincidence problem for the cosmological constant, although it does relax the former. These issues will require a better understanding of the regularization mechanism (see below), which goes beyond the scope of this work.}

This is achieved by extending the analysis of \cite{AB1}, which was applied near a cosmological bounce where the gap equation could be obtained in a Minkowski spacetime. Here we work on a FLRW background and consider the evolution of the universe from the bounce on. As a consistency check, we obtain gap equations similar\footnote{Some of the coefficients differ slightly. This is mainly because here for simplicity we only consider the scalar channel of the four-fermion interaction, while in~\cite{AB1} we also included the pseudoscalar interaction. Also, unlike in~\cite{AB1}, we do not throw away the contributions from the antifermions. These differences do not affect the physics, since the coefficients have the same sign.} to those in~\cite{AB1} and reproduce the same bouncing cosmology. A mechanism of fermion condensation was argued to be relevant to dark energy also in \cite{CaC,IMM} (see also \cite{AC1,AC2}); for another approach describing fermions and condensates on FLRW spacetimes, see \cite{CR,GHN,KoP}.

The paper is organized as follows. In Sec.~\ref{BCS} we introduce the BCS mechanism on a FLRW background and derive the effective equations of motion. This section is a little technical and the reader mostly interested in the cosmology can skip ahead to Sec.~\ref{sol}, where we classify the cosmological solutions and find  numerical examples which accelerate at late times. Section \ref{disc} is devoted to discussion.


\section{Cosmological BCS theory}\label{BCS}

We use Greek indices $\mu, \nu, \dots =0,\dots, 3$ for spacetime directions in a nondegenerate manifold ${\cal M}$ (with signature ${-}{+}{+}{+}$)
and capital Latin indices $I, J, \dots =0,\dots, 3$ for the internal Lorentzian tangent space $T_{\cal M}$. Spatial directions on ${\cal M}$ and $T_{\cal M}$ will be denoted as $a,b,c,\dots$ and $i,j,k,\dots$, respectively. Dotted and undotted indices $\dot\al, \al$ label spinor components. We work in units $\hbar=c=G=1$.

We begin by studying the cosmology of first-order general relativity covariantly coupled to Dirac fermions. In such a system a four-fermion interaction emerges when we solve for the torsion. In what follows, we show that this interaction in a FLRW background realizes a BCS condensate whose potential generically leads to late-time acceleration.


\subsection{Fermions in FLRW background}

Let us start by considering pure general relativity described by the Holst action \cite{Hol95}. Afterwards we sketch how the attractive four-fermion interaction emerges after solving for the torsion. The action on a curved manifold $\cM$ is
\be\label{faction}
S_{\rm H} = \frac{1}{16\pi} \left(\int_\cM\rmd^4x\,e\,e_I^\mu\,e_J^\nu\,R_{\mu\nu}^{\ \ IJ}
   - \frac{1}{\gamma} \int_\cM \rmd^4x\,e\,e_I^\mu\,e_J^\nu\,\tilde{R}_{\mu\nu}^{\ \ IJ} \right)\,,
\ee
where $e_\mu^I$ is the gravitational field (vielbein, tetrad), $e\equiv |\det e^\mu_I|$, $R_{\mu\nu}^{\ \ IJ}$ is the curvature of the spin connection $A^{\ IJ}_\mu$, $\gamma$ is the Barbero--Immirzi parameter and
$\tilde{R}_{\mu\nu}^{\ \ IJ}=\epsilon^{IJ}_{\ \ KL}\,R_{\mu\nu}^{\ \ KL}/2$ is the dual field strength.

The first term in Eq.~\Eq{faction} yields the Palatini formulation of the Einstein--Hilbert action, the latter emerging when inserting the solution to the associated equation of motion
\be\label{conn}
A_{\ IJ}^\mu=\omega_{\ IJ}^\mu[e] = e_I^\nu \left(e_{J,\nu}^\mu - \Gamma_{\rho\nu }^\mu\,e_J^\rho\right)
\ee
(where $\Gamma_{\rho\nu}^\mu$ are the Christoffel symbols) into Eq.~\Eq{faction} and using $g_{\mu\nu}=e_\mu^I\,e_{\nu}^J \eta_{IJ}$. The second term is identically zero on half-shell, due to the Bianchi identity for the Riemann tensor. $\omega^{\ KL}_\mu$ is the metric-compatible, torsion-free spin connection.

The story changes when we covariantly couple chiral fermions to the gravitational action. The Dirac action for fermions $\psi$ is
\be\label{actd}
S_{\rm D} =- \frac{\rmi}{2} \int_{\cal M} \rmd^4x e \left(\bar{\psi} \gamma^I e_I{}^\mu \n_\mu \psi + {\textrm{c.c.}}\right)\,,
\ee
where c.c. denotes the complex conjugate and the covariant derivative is defined as
$\n_\mu = \partial_\mu -\Oa_\mu$, where
\be
\Oa_\mu=  \frac{1}{4} A_{IJ\mu} \Sigma^{IJ}\,,
\ee
and
\be
\Sigma^{IJ}\equiv\2[\gamma^I,\gamma^J]
\ee
are the Lorentz generators for spinors. $\gamma^I$ are the usual Dirac matrices in Weyl (chiral) basis,
\be
\gamma^I = \begin{pmatrix} 0 & \sigma^I \\ \bar\sigma^I & 0 \end{pmatrix},\qquad \gamma_5 = \begin{pmatrix} \mathbb{I} & 0 \\ 0 & -\mathbb{I} \end{pmatrix},
\ee
where $\sigma^I=(\mathbb{I},-\sigma^i)$, $\bar\sigma^I=(\mathbb{I},\sigma^i)$ and $\sigma^i$ are the Pauli matrices:
\be
\sigma^1 =
\LF\begin{array}{cc} 0&1\\
1&0
\end{array}\RF\,,\qquad \sigma^2 =
\LF\begin{array}{cc} 0&-\rmi\\
\rmi&0
\end{array}\RF\,,\qquad \sigma^3 =
\LF\begin{array}{cc} 1&0\\
0&-1
\end{array}\RF\,.
\ee
Note that a tetrad-based formalism is essential for the inclusion of fermions in the theory, since Dirac spinors live naturally in $SU(2)$.

The equation of motion for the total action $S_{\rm H}+S_{\rm D}$ are solved in terms of a connection $A^{\ IJ}_\mu$ having two contributions \cite{PR,FMT,Mer06}, the torsion-free spin connection and a torsion term related to the axial fermion current:
\be\label{connection}
A_{\mu}^{\ IJ}=\omega_{\mu}^{\ IJ}[e] + C_{\mu}^{\ IJ}\,,
\ee
where $C_{\mu}^{\ IJ}$ is the tetrad projection of the contortion tensor,
\be\label{contortion}
C_{\mu}^{\ IJ} = C_{\mu}^{\ \nu\delta}e^{I}_{[\nu}e^{J}_{\delta]}\,.
\ee
Square brackets denote antisymmetrization, $X_{[IJ]}=(X_{IJ}-X_{JI})/2$.

On solving for $C_{\mu}^{\ IJ}$ in terms of the fermionic field and inserting the resulting
expression for $A_{\mu}^{\ IJ}$ in the total action, one obtains the four-fermion interaction (see the Appendix for a detailed derivation)
\be
S_{\rm int}=\int_{\cal M} \rmd^4x\, e\frac{J_{5I} J^I_5}{M^2}\,,\qquad \frac{1}{M^2}=\frac{3\pi}{2}\frac{\gamma^2}{\gamma^2+1}\,,\label{interaction}
\ee
where $J_5^I=\bar\psi\gamma_5\gamma^I\psi$ is the axial current. Therefore, the fermionic action is now
\be\label{sfer}
S_{\rm fer}=S_{\rm D}+S_{\rm int}\,,
\ee
where the covariant derivative in $S_{\rm D}$ is in terms of ${\oa}^{\ IJ}_\mu$. For a flat FLRW line element
\be
\rmd s^2=-\rmd t^2+a^2(t)\,\rmd x_a \rmd x^a\,,
\ee
the vielbein is given by
\be
e_\mu{}^I=\LF\begin{array}{cc}
1 &0\\
0& a(t)\da_{a}^{i}
\end{array}\RF\,,
\ee
where $a(t)$ is the scale factor. The only nonzero structure functions and spin connections $\oa_{IJK}$ are
\be\label{CC}
\oa_{0\au\bu}=-\oa_{\au0\bu}=-H\delta_{\au\bu}\,,
\ee
where $H\equiv \dot a/a$ is the Hubble parameter. Then,
\be
\Oa_0=0\,,\qquad \Oa_{a}=-\frac{aH}{2}\delta_{a\bu}\LF\begin{array}{cc}
\sa^{\bu} &0\\
0& -\sa^{\bu}
\end{array}\RF\,.
\ee
Above we have assumed the gravitational action to be the usual Holst action. However, when torsion is present (for instance, when it is generated by fermions or a spacetime-dependent Barbero--Immirzi field \cite{TY,TK,CM,Mer09,MT}), it is natural to include it explicitly in the fundamental action, so that the Holst term is completed by a torsion-torsion piece to form the Nieh--Yan invariant \cite{Mer06,Mer08}.

There are several other reasons why to prefer the latter alternative. A second motivation is that the Holst term is not topological and vanishes on half-shell, while one would expect to define the theory with topological contributions. Third, although mathematically correct the Holst derivation does not respect the usual decomposition of torsion into its Lorentz irreducible components, and its trace part (a polar internal vector) turns out to be proportional to the axial current \cite{Mer06} [see Eq.~\Eq{vecax}].

In the Nieh--Yan case, the coupling $M^2=2/(3\pi)$ no longer depends on the Barbero--Immirzi parameter. In either case the ``bare'' coupling is $M\gtrsim 1$. As we will regard it as part of an effective coupling, our results will not be sensitive to the form of the classically vanishing part of the action.


\subsection{Weyl decomposition}

We are now ready to quantize the fermions on a FLRW spacetime. Since
\ba
e_I{}^\mu\ga^I\n_\mu\psi&=&\LT\ga^0\p_0+\frac{1}{a}\ga^{\au}\da_{\au}^{a}(\p_{a}-\Oa_{a})\RT\psi\nonumber\\
&=&\ga^0\LF\dot{\psi}+\frac{3}{2}H\psi\RF+\frac{1}{a}\ga^{\au}\p_{\au}\psi\,,\nonumber
\ea
the Dirac Lagrangian is
\be
\cL_{\rm D}=-\rmi\psi\+\ga^0\LT\ga^0\LF\dot{\psi}+\frac{3}{2}H\psi\RF+\frac{1}{a}\ga^{\au}\p_{\au}\psi\RT\,.
\ee
Just as in the case of Minkowski spacetime, it is convenient to decompose the fermions into two-component Weyl spinors,
\be
\psi\equiv \LF\begin{array}{c} \xi\\ \chi\end{array}\RF\,,
\ee
so that
\be\label{dirla}
\cL_{\rm D}=-\rmi\LT\xi\+\dot{\xi}+\chi\+\dot{\chi}+\frac{1}{a}\LF\xi\+\se^{\au}\p_{\au}\xi+\chi\+\sa^{\au}\p_{\au}\chi\RF+\frac{3}{2}H\LF\xi\+\xi+\chi\+\chi\RF\RT\,.
\ee
One can also write the action in terms of left-handed Weyl spinors but there is now an extra term coming from the integration by parts (integration domain omitted from now on):
\be
\int \rmd^4x\ a^3\chi\+\dot{\chi}=\int \rmd^4x\ \chi\p_t(a^3\chi\+)=\int \rmd^4x\ a^3 \chi(\dot{\chi}\++3H\chi\+)\,,\nonumber
\ee
since the $\chi$'s are anticommuting Grassmann fields, $\chi\+\chi=-\chi\chi^{\dagger}$. For spatial derivatives,
\be
\int \rmd^4x\ a^2\chi\+\sa^{\au}\p_{\au}\chi=-\int \rmd^4x\ a^2\p_{\au}\chi\se^{\au}\chi\+=\int \rmd^4x\ a^2\chi\se^{\au}\p_{\au}\chi\+\,,\nonumber
\ee
and Eq.~\Eq{dirla} yields
\be
S_{\rm D}
=-\rmi\int \rmd^4x\ a^3\LT \xi\+\dot{\xi}+\za\+\dot{\za}+\frac{1}{a}\LF\xi\+\se^{\au}\p_{\au}\xi+\za\+\se^{\au}\p_{\au}\za\RF+\frac{3}{2}H\LF\xi\+\xi+\za\+\za\RF\RT\,,
\ee
where $\za=\chi\+$. The action is completely symmetric with respect to the particle and antiparticle Weyl spinors, $\xi\leftrightarrow\za$. In other words, the expansion of the universe does not distinguish between particles and antiparticles.

Moreover, since FLRW is conformally flat and the action is first-order in time derivatives, the latter reduces to the Minkowski action. To see this, let us first perform the conformal rescaling
\be
\w{\xi}= a^{3/2}\xi\,,\qquad \w{\za}= a^{3/2}\za\,.
\ee
Then,
\be
S_{\rm D}=-\rmi\int \rmd^4x \LT\w{\xi}\+\dot{\w{\xi}}+\w{\za}\+\dot{\w{\za}}+\frac{1}{a}\LF\w{\xi}\+\se^{\au}\p_{\au}\w{\xi}+\w{\za}\+\se^{\au}\p_{\au}\w{\za}\RF\RT\,.
\ee
Next we introduce the Fourier transforms
\be
\w{\xi}(x)=\int \rmd^3\vec{p}\,\rmd\oa\ \rme^{-\rmi[a\vec{p}\cdot\vec{x}-\oa t]}\w{\xi}_{\vec{p},\oa}\,,\qquad \w{\za}(x)=\int \rmd^3\vec{p}\,\rmd\oa\ \rme^{-\rmi[a\vec{p}\cdot\vec{x}-\oa t]}\w{\za}_{\vec{p},\oa}\,.
\ee
In terms of the Fourier components, the action becomes
\ba
S_{\rm D}&=&\int \rmd^4p\,\rmd^4p'\,\rmd^4x \LT\oa\w{\xi}_{\vec{p}',\oa'}\+\w{\xi}_{\vec{p},\oa}+\oa\w{\za}_{\vec{p}',\oa'}\+\w{\za}_{\vec{p},\oa}-\w{\xi}_{\vec{p}',\oa'}\+\se^{\au}p_{\au}\w{\xi}_{\vec{p},\oa}
-\w{\za}_{\vec{p}',\oa'}\+\se^{\au}p_{\au}\w{\za}_{\vec{p},\oa}\RT\nonumber\\
&&\qquad\qquad\qquad\times \rme^{-\rmi[a\vec{x}\cdot(\vec{p}-\vec{p}')+t(\oa-\oa')]}\nonumber\\
&=&\int \rmd^4p\,\rmd \oa'\,\rmd t \left(\frac{2\pi}{a}\right)^3\LT\oa\w{\xi}_{\vec{p},\oa'}\+\w{\xi}_{\vec{p},\oa}+\oa\w{\za}_{\vec{p},\oa'}\+\w{\za}_{\vec{p},\oa}-\w{\xi}_{\vec{p},\oa'}\+\se^{\au}p_{\au}\w{\xi}_{\vec{p},\oa}
-\w{\za}_{\vec{p},\oa'}\+\se^{\au}p_{\au}\w{\za}_{\vec{p},\oa}\RT\nonumber\\
&&\qquad\qquad\qquad\times  \rme^{-\rmi t(\oa-\oa')}\nonumber\\
&=&(2\pi)^3\int \rmd^4p\,\rmd\oa'\rmd t\LT\oa\xi_{\vec{p},\oa'}\+\xi_{\vec{p},\oa}+\oa\za_{\vec{p},\oa'}\+\za_{\vec{p},\oa}
-\xi_{\vec{p},\oa'}\+\se^{\au}p_{\au}\xi_{\vec{p},\oa}
-\za_{\vec{p},\oa'}\+\se^{\au}p_{\au}\za_{\vec{p},\oa}\RT\nonumber\\
&&\qquad\qquad\qquad\times \rme^{-\rmi t(\oa-\oa')}\,.\nonumber
\ea
Thus, the Dirac action in momentum space reads
\be\label{dirp}
S_{\rm D}=(2\pi)^4\int \rmd^4p\LT\oa(\xi_{\vec{p},\oa}\+\xi_{\vec{p},\oa}+\za_{\vec{p},\oa}\+\za_{\vec{p},\oa})
-(\xi_{\vec{p},\oa}\+\se^{\au}p_{\au}\xi_{\vec{p},\oa}
+\za_{\vec{p},\oa}\+\se^{\au}p_{\au}\za_{\vec{p},\oa})\RT\,.
\ee


\subsection{BCS condensation}

A simple and physically transparent way to understand the condensation mechanism is to introduce auxiliary scalar (gap)
fields, which are proportional to the fermionic bilinears. The gap equation is then derived by
integrating out the fundamental fermionic degrees of freedom. Our starting point is the four-fermion interaction term
\be
S_{\rm int}=\int \rmd^4x e\LT \frac{J_{5I} J^I_5}{M^2}\RT\,. \label{action}
\ee
This term can be decomposed into a scalar, pseudoscalar and vector interactions using the Fierz identity
\be
(\bar{\psi} \gamma_5 \gamma^I \psi)(\bar{\psi} \gamma_5 \gamma_I \psi)=(\bar{\psi}  \psi)^2+(\bar{\psi} \gamma_5\psi)^2+(\bar{\psi} \gamma^I \psi)(\bar{\psi} \gamma_I \psi)\,.
\ee
The last term is the higher-energy $p$-wave channel and we, as such, are going to ignore it. For simplicity, we will also drop the pseudoscalar condensate and only focus on the scalar one. Thus, our interaction reduces to
\be
S_{\rm int}=\int \rmd^4x\, e\LT \frac{(\bar{\psi}\psi)^2}{M^2}\RT =\int \rmd^4x\, e\LT (\bar{\psi}\psi)\Da-\frac{M^2}{4} \Da^2\RT\equiv S_{\rm mass}+S_{\rm tree}\,,
\ee
where in the second equality we have introduced the auxiliary scalar $\Da$, which acts like a mass term for the fermions. For an FLRW background (spinorial indices restored),
\be
S_{\rm mass}=\int \rmd^4x\, e (\bar{\psi}\psi)\Da=\int \rmd^4x\, a^3(\en^{\al\beta}\za_{\beta}\xi_{\al}+\en^{\ad\bd}\xi\+_{\ad}\za\+_{\bd})\Da\,.
\ee
It is clear that a nonzero value for the auxiliary field $\Da\sim\bar{\psi}\psi$ would signal a (cosmological) BCS-like condensation. In order to find such a nontrivial value for $\Da$, one can take recourse to a mean-field approximation where the gap $\Da$ is treated as a constant. With the same procedure of the last subsection, the mass term in momentum space is
\be
S_{\rm mass}=(2\pi)^4\int \rmd^4p(\en^{\al\beta}\za_{\beta,-\pv,-\oa}\xi_{\al,\pv,\oa}+\en^{\ad\bd}\xi\+_{\ad,\pv,\oa}\za\+_{\bd,-\pv,-\oa})\Da\,.
\ee
Combining it with the kinetic term \Eq{dirp},
\ba
S_{\rm fer}&\equiv& S_{\rm D}+S_{\rm mass}\nonumber\\
&=&(2\pi)^4\int \rmd^4p\LT\oa\xi_{\vec{p},\oa}\+\xi_{\vec{p},\oa}-\xi_{\vec{p},\oa}\+\se^{\au}p_{\au}\xi_{\vec{p},\oa}
+\oa\za_{-\vec{p},-\oa}\za_{-\vec{p},-\oa}\+-\za_{-\vec{p},-\oa}\sa^{\au}p_{\au}\za_{-\vec{p},-\oa}\+\right.\nonumber\\
&&\qquad\qquad\qquad\qquad+\left.(\za_{-\pv,-\oa}\xi_{\pv,\oa}+\xi\+_{\pv,\oa}\za\+_{-\pv,-\oa})\Da\RT\,,
\ea
where $\xi=\xi_{\al}$ and $\za=\za^{\al}$. The above can be written in four-component notation as
\be
S_{\rm fer}=(2\pi)^4\int \rmd^4p\ (\xi_{\vec{p},\oa}\+,\za_{-\pv,-\oa})A_p
\LF\begin{array}{c}
\xi_{\pv,\oa}\\ \za\+_{-\pv,-\oa}
\end{array}\RF\,,
\ee
where $A_p$ is a $4\times 4$ matrix given by
\be
A_p=\LF\begin{array}{cc} \oa-\se^{\au}p_{\au}&
\Da\\
\Da& \oa-\sa^{\au}p_{\au}
\end{array}\RF\,.
\ee

At this point we introduce a chemical potential $\mu$ in the action, which corresponds to having a nonzero number density of fermions. The matrix $A_p$ is now modified to
\be
A_p=\LF\begin{array}{cc} \oa-\se^{\au}p_{\au}+\mu&
\Da\\
\Da& \oa-\sa^{\au}p_{\au}-\mu
\end{array}\RF\,.
\ee
The condition $\mu<0$ corresponds to a Bose--Einstein condensation of composite bosons.


\subsection{Effective action}\label{epl}

The resulting quantum theory is encoded into the path integral
\be\label{effa}
Z=\int [\cD\Da][\cD\xi][\cD\za]\rme^{\rmi(S_{\rm fer}+S_{\rm tree})}
\equiv \int [\cD\Da]\rme^{\rmi S_{\rm eff}}\approx \rme^{\rmi S_{\rm eff}}\big|_{\rm SP}\,,
\ee
where we have integrated the Grassmann fields, defined the effective action $S_{\rm eff}$ (often referred to as $\Ga$ in quantum field theory literature) and approximated the functional integral by the saddle point (mean-field approximation \cite{Sch99}). The effective action $S_{\rm eff}$ can be evaluated by performing the Gaussian integrals in terms of the fermionic coordinates. As usual, one ends up with a fermionic determinant. Eventually we have (see, e.g., \cite{Sch99,KoS})
\be
S_{\rm eff}=S_{\rm tree}-\rmi\int \frac{\rmd^4p}{(2\pi)^4} \ln (\det A_p)\,.
\ee
The determinant of $A_p$ can be straightforwardly computed:
\ba
\det A_p=[\oa^2-(|\pv|+\mu)^2-\Da^2][\oa^2-(|\pv|-\mu)^2-\Da^2]\,,
\ea
where $\Delta$ is the auxiliary field at the saddle point. This expression is not Lorentz covariant, as expected (see below). Accordingly, we are left computing
\be
V_{\rm eff}\equiv -\cL_{\rm eff}=\frac{M^2}{4}\Da^2-I\,,
\ee
where
\ba
I &=&\int \frac{\rmd^3\vec{p}\rmd \oa}{(2\pi)^4}\ \{\ln[\oa^2-(|\pv|+\mu)^2-\Da^2]+\ln[\oa^2-(|\pv|-\mu)^2-\Da^2]\}\nonumber\\
&=& \int \frac{\rmd^3\vec{p}}{(2\pi)^3}\ \left[\sqrt{(|\pv|+\mu)^2+\Da^2}+\sqrt{(|\pv|-\mu)^2+\Da^2}\right]\nonumber\\
&=& I_1+I_2\,.
\ea
The integral $I_1$ is
\ba
I_1&=&\int_0^{\infty} \frac{\rmd p}{2\pi^2}\ p^2\sqrt{(p+\mu)^2+\Da^2}=\int_{\mu}^{\infty} \frac{\rmd p}{2\pi^2}\ (p-\mu)^2\sqrt{p^2+\Da^2}\nonumber\\
&=&\int_{0}^{\infty} \frac{\rmd p}{2\pi^2}\ (p-\mu)^2\sqrt{p^2+\Da^2}-\int_{0}^{\mu} \frac{\rmd p}{2\pi^2}\ (p-\mu)^2\sqrt{p^2+\Da^2}\,.\nonumber
\ea
To get $I_2$, one has to simply replace $\mu\leftrightarrow-\mu$ in $I_1$.
The second integral in $I_1$ cancels the second in $I_2$ and we are left with
\be
I=\int_{0}^{\infty} \frac{\rmd p}{\pi^2}\ (p^2+\mu^2)\sqrt{p^2+\Da^2}\,.
\ee
The above integral can be regulated using the following formula:
\be
\int_0^{\infty}\rmd p\ \frac{p^A}{(p^2+\Delta^2)^B}=\frac{\Ga\LF\frac{1+A}{2}\RF\Ga\LF B-\frac{1+A}{2}\RF}{2\Delta^{2B-A-1}\Ga(B)}\,.
\ee
Let us choose $B=-1/2+\en$. Then we have ($A=2$ and $A=0$)
\ba
I&=&\frac{1}{2\pi^2}\LT\frac{\Ga(3/2)\Ga(\en-2)}{\Da^{2(\en-2)}\Ga(\en-1/2)}+\frac{\mu^2\Ga(1/2)\Ga(\en-1)}{\Da^{2(\en-1)}\Ga(\en-1/2)}\RT\nonumber\\
&=&\frac{\sqrt{\pi}\Da^2}{2\pi^2\Ga(\en-1/2)\Da^{2\en}}\LT\2\Da^2\Ga(\en-2)+\mu^2\Ga(\en-1)\RT\,.\nonumber
\ea
We expand up to $\cO(\en)$ to obtain the effective action. Using expansion formul\ae\ such as
\be
\Ga\LF\en-\2\RF=-2\,\sqrt {\pi }+2\,\left(\gamma-2+2\,\ln 2\right) \sqrt {\pi }\en+\cO(\en^2)\,,
\ee
where $\ga\approx 0.5772$ is the Euler--Mascheroni constant, we have
\ba
I&\approx&-\frac{\Da^2}{4\pi^2[1-\en(\ga-2+2\ln 2)](1+\en\ln\Da^2)}\LT\frac{\Da^2}{4}\LF \frac{1}{\en}-\ga+\frac{3}{2}\RF-\mu^2\LF\frac{1}{\en}-\ga+1 \RF\RT\nonumber\\
&\approx& -\frac{\Da^2}{4\pi^2}\LT\4\Da^2\LF \frac{1}{\en}-\2+2\ln 2-\ln\Da^2\RF-\mu^2\LF\frac{1}{\en}-1+2\ln 2-\ln\Da^2 \RF\RT\,.
\ea
Thus, the effective potential is given by
\be
V_{\rm eff}=\frac{M^2}{4}\Da^2+\frac{\Da^2}{4\pi^2}\LT\frac{\Da^2}{4}\LF \frac{1}{\en}-\2+2\ln 2-\ln\Da^2\RF-\mu^2\LF\frac{1}{\en}-1+2\ln 2-\ln\Da^2\RF\RT.
\ee
In a renormalizable theory, the $1/\en$ divergence can be absorbed using renormalization conditions \cite{ELOS,IMO}. The four-fermion interaction term is nonrenormalizable in Minkowski and therefore the divergence cannot be eliminated. A standard approach is to interpret the regularization parameter in terms of a physical cutoff scale $\la$,
\be\label{lamb}
\frac{1}{\en}\sim\ln\la^2\,,
\ee
such that $\la$ remains \emph{finite}. Here we just take a phenomenological approach and  encode this arbitrariness, intrinsic to the model, in a reparametrization of the form
$\la=\rme^{-N/2}/2$, where $N$ is an $\cO(1)\div\cO(10^2)$ free parameter:
\be\label{en}
\frac{1}{\en}=-N-2\ln2\,.
\ee
The relation between $\en$ and $\la$ is really a matter of choice, so any sign and value of $N$ is possible. If $N>0$ then $\en<0$, which may happen if spacetime has a fractal structure in the ultraviolet. Later we will need $N\gtrsim \cO(10^2)$ in order for the condensate to fit observations. Interestingly enough, such a range of values $-1\ll\en<0$ may be compatible with a fractal interpretation of $N$, where the (early) universe shows an effective dimension slightly smaller than 4. Until we achieve a better control of the quantum theory, the issue of the physical interpretation of $N$ will remain open, although we have just argued that it admits at least one possible accommodation.

To summarize, the effective potential is
\be\label{Veff}
V_{\rm eff}=\frac{M^2}{4}\Da^2-\frac{\Da^2}{4\pi^2}\LT\frac{\Da^2}{4}\LF N+\2+\ln\Da^2\RF-\mu^2\LF N+1+\ln\Da^2\RF\RT\,.
\ee
As already mentioned, the bare mass $M$ is $\cO(1)$ or larger. However, Eq.~\Eq{Veff} can be obtained also via a different regularization procedure which renormalizes the couplings of the theory \cite{AB1,Sch99}. This suggests that also $M$ can be treated as a free parameter. It shall be our attitude in what follows.

It is clear that the above potential has a minimum at $\p V_{\rm eff}/\p \Da=0$ given by the \emph{gap equation}
\be
M^2=\frac{1}{2\pi^2}\LT\Da^2(N+1)-2\mu^2(N+2)+\LF\Da^2-2\mu^2\RF\ln\Da^2 \RT. \label{gap}
\ee
If $M$ is constant, this equation univocally specifies $\Da$ as a spacetime function. It is useful to check that we recover the usual behaviour of the gap in the weak-coupling BCS limit~\cite{Sch99,KoS}, where the fermion gas is diluted. For $\Da\ll\mu$, Eq.~\Eq{gap} tells us that
\be
\Da\approx \exp\LF-\frac{\pi^2M^2}{2\mu^2}\RF \label{exp-gap}\,,
\ee
which is the familiar exponential suppression of the gap.

The potential at the minimum is
\be
V_{\rm min}=\frac{\Da^2}{16\pi^2}\LT\Da^2\LF N+\frac{3}{2}+\ln\Da^2\RF-4\mu^2\RT\,.
\ee
However, the potential energy that we have calculated includes the contribution from the chemical potential as well. The total number $n_0$ of fermions is~\cite{Sch99}
\be\label{neq}
n_0=\int \rmd^4 x \ e\bar{\psi}\ga^0\psi=\frac{\delta S_{\rm fer}}{\delta \mu}=-a^3 \frac{\p V_{\rm eff}}{\p \mu}=-a^3\frac{\Da^2\mu}{2\pi^2}\LF N+1+\ln\Da^2\RF\equiv a^3 n\,.
\ee
Assuming the system lies at the minimum of the potential, the total gap energy density of the fluid is given by
\ba
\rho_{\rm gap}&=&V_{\rm min}+\mu n\nonumber\\
&=& \frac{\Da^2}{16\pi^2}\LT\Da^2\LF N+\frac{3}{2}+\ln\Da^2\RF-4\mu^2(2N+3+2\ln\Da^2)\RT\nonumber\\
&=& \frac{\Da^2}{32\pi^2}\LF\Da^2-8\mu^2\RF \LF 2N+3+2\ln\Da^2\RF\,.\label{rga}
\ea

We conclude this section by making two remarks, the first on the chemical potential. As soon as we fixed an FLRW background, we have chosen a ``privileged'' frame whereon one can define a homogeneous number density and a chemical potential. Whenever we refer to these concepts, it is always with respect to this special FLRW homogeneous time slicing. Solutions of the field equations are free to break Lorentz invariance. One can have a chemical potential in a Lorentz-invariant Lagrangian, i.e., a term of the form $\mu_{\nu}j^{\nu}$, while the background $\mu_{0}j^{0}$ is aligned in the cosmic rest frame. An example is the Kaplan--Nelson model of spontaneous baryogenesis. This is what happens also in big bang nucleosynthesis calculations, where one has to include the chemical potential of the different particle species.\footnote{In this context we note that having a nonzero chemical potential implies that we have broken the particle-antiparticle symmetry, which in turn breaks local Lorentz invariance regardless of the chosen global metric. Thus we are working on the assumption that some other mechanism was responsible for creating the ``initial'' particle-antiparticle asymmetry.}

This has nothing to do with breaking Lorentz invariance at fundamental level. Here we are assuming, in a self-consistent manner, that the theory and its false (perturbative) vacuum are indeed Lorentz-invariant, while the true vacuum is not when $\mu\neq 0$. We will see that the measurable effect is in the cosmic expansion, and the later one looks into the evolution of the universe, the fewer the Cooper pairs one can detect, thus restoring relativistic physics at late times.

The second comment is the following. We have introduced a physical cutoff by hand as in Minkowski nonrenormalizable theories, but in \cite{GHN} it was argued that this type of BCS models in curved space may be renormalizable. In the same work, the running of the mass coupling $M$ and a cosmological constant term was considered for a de Sitter background. Here we shall not endeavour to study the renormalization group flow for the particular, self-consistent cosmological background we will find later. However, related to the renormalization issue there is another. Although the tree-level interaction between fermions is attractive, the gravitational interaction could affect renormalization. This is in analogy with the phonon interaction in the standard theory, where the sign of the effective coupling gets flipped and one does not generically end up with a condensate \cite{Pol92,Sha94}. There, one can consider the beta function for the running of the coupling $M$, and realize that the second-order (four-fermion) phonon-induced interaction is attractive (negative coupling) \cite{Pol92}. In our case, the graviton and the torsion are independent degrees of freedom and they can be dealt with separately. An attractive fermion coupling is generated by torsion after integrating it out. We may ask about torsion fluctuations, but they do not take place because torsion is not dynamical. Therefore, so long as we have an attractive channel at tree level and a Fermi surface, we will get a nontrivial gap (provided that the gap equation has a solution, which it does; see below). As far as torsion is concerned, the attractive nature of the tree-level interaction is guaranteed, unlike the phonon-induced screening (or gluon exchange) in typical scenarios. In our case, we do not have to worry about a secondary phonon interaction precisely because torsion does not propagate. On the other hand, any screening effect mediated by the graviton will require a separate, detailed assessment.


\section{Cosmological solutions}\label{sol}

In order to obtain the cosmological evolution one has to supplement the gap equation (\ref{gap}) and the equation determining the chemical potential (\ref{neq}) with the Friedmann equation
\be\label{freq}
H^2=\frac{8\pi}{3}(\rho_{\rm gap}+\rho_{\rm m})\,,
\ee
where $\rho_{\rm m}$ is any additional matter component. These three expressions (plus the continuity equation for $\rho_{\rm m}$) determine the evolution of the three unknown quantities $\Da(t)$, $\mu(t)$, $a(t)$. In practice, these transcendental equations are not analytically tractable in their full generality. However, one can solve them in some limiting regimes which can qualitatively capture the basic cosmological evolution. We will now focus on these regimes. In the next subsection we will present numerical details.

The equation of state for the gap can be defined only implicitly via Eqs.~\Eq{gap} and \Eq{rga}. We will show that there are different regimes where the gap scales effectively as radiation (up to a sign) or a cosmological constant, which confirm the expectation that the gap equation of state must be nonlinear. One can define the gap pressure by taking an effective Raychaudhury equation, i.e., by defining the effective barotropic index via 
\be\label{weff}
1+w_{\rm eff}=-\frac13\frac{d\ln|\rho_{\rm gap}|}{d\ln a}\,,
\ee
as a function of the scale factor. This is what dark energy experiments would measure.


\subsection{Analytical treatment}

\subsubsection{Dark radiation (early times)}

To begin with, it is important to see whether the evolution of the gap and the chemical potential at early times can be consistent with the usual matter/radiation dominated decelerating universe. To answer this question, let us look into the limiting case $M\ll|\mu|,\Da$ and $N\gg 1$ (later we motivate the last condition phenomenologically). The gap equation simplifies to
\be
\Da^2\approx 2\mu^2\,.
\ee
The energy density of the gap \Eq{rga} becomes
\be
\rho_{\rm gap}\approx -\frac{3\Da^4}{32\pi^2} \LF 2N+3+2\ln\Da^2\RF<0\,.
\ee
One can also determine the approximate behaviour of the energy density with respect to the scale factor. From Eq.~(\ref{neq}), we find that $\mu$ must be negative and
\be
\frac{\Da^3}{2\stwo\pi^2}\LF N+1+\ln\Da^2\RF= \frac{n_0}{a^3}\ ,
\ee
or approximately (if $\Delta$ varies slowly)
\be
\Da\sim \frac{1}{a}\qquad\Ra\qquad \rho_{\rm gap}\sim -\frac{1}{a^4}\,.
\ee
In other words, the gap energy density behaves approximately as negative radiation.\footnote{A similar contribution appears in braneworld \cite{BDEL} and Ho\v{r}ava--Lifshitz \cite{Lif1} cosmologies.} This fact was already pointed out in~\cite{AB1}. It is also clear that the gap energy density violates the null energy condition\footnote{This is not very surprising, as it is well known that energy densities associated  with vacuum shifts, such as Casimir energies, can indeed violate the energy conditions~\cite{casimir}.} and this property was exploited in~\cite{AB1} to resolve the big bang singularity via a nonsingular bounce, as long as the equation of state of ordinary matter is $w<1/3$.

\subsubsection{de Sitter phase (late times)}

It is relatively easy to see how a late-time de Sitter phase can emerge from the system of cosmological equations. If $|\mu|\ll\Da$, the gap equation reduces to ($\mu<0$ in order to have $n>0$)
\be
2\pi^2M^2\approx  \Da^2\LF N+1+\ln\Da^2\RF\,,\label{gap-dS}
\ee
while the total energy density is given by
\be\label{boh}
\rho_{\rm gap}\approx \frac{\Da^4}{32\pi^2} \LF 2N+3+2\ln\Da^2\RF\,.
\ee
First, we observe that the solution to the above equation always has $\rho_{\rm gap}>0$, and therefore corresponds to a de Sitter regime as $\Da$ approaches the constant value given by Eq.~\Eq{gap-dS}. This can be seen by rewriting Eq.~\Eq{boh} as
$$
\rho_{\rm gap}\approx\frac{\Da^4}{32\pi^2} \LF \frac{4\pi^2M^2}{\Da^2}+1\RF>0\,.
$$

We are specifically interested to see whether we can explain the present dark energy driven acceleration. For this we require (we temporarily restore energy units) $\rho_{\rm gap}\sim ({\rm meV})^4\Ra \Delta\sim {\rm meV}$. When can we have such small vacuum expectation value for $\Da$? First of all $\Da \sim {\rm meV}$ corresponds to $\ln \Da^2\approx -140$. Since $M^2>0$, (\ref{gap-dS}) tells us that $N> 140$. Further, from (\ref{gap-dS}) it is easy to see that there are two different regimes in the parameter space $(M,N)$ when we can get a small vacuum expectation value for $\Da$. If $M\sim {\rm meV}$ and $140<N<10^3$, the solution corresponds to $\Da\sim M\sim {\rm meV}$. A second possibility is to consider $M\ll {\rm meV}$ and $N\sim 140$. In this case we have $\Da\sim e^{-N/2}\sim {\rm meV}$. In any case, the relevant range of parameters corresponds to large $N$ and strong-coupling regime,
\be
N\gg 1\,,\qquad M\ll 1\,.
\ee
At this point one may be concerned about the tiny value of $M$ that is required to account for dark energy. Indeed, naturalness arguments would suggest $M\sim 1$ (Planck scale). We first point out that the $M$ appearing in the effective potential \Eq{Veff} can be interpreted as a renormalized mass (see~\cite{AB1} for a more detailed discussion), and therefore in general it can be different from the bare coupling mass in Eq.~\Eq{interaction}. We shall not study the renormalization group flow of this model, and the fact that it is an effective nonrenormalizable model makes it rather difficult to interpret the relation between physical scales and parameters. A much deeper understanding of the quantum field theory giving rise to the condensate will be necessary to clarify this point, which will admittedly remain unresolved in this paper. 

We also mention that there are two other possible explanations for a small coupling $M$. One is tightly related with the fractal interpretation given below Eq.~\Eq{en}. Another is to switch on a strong four-fermion interaction already in the Dirac action \Eq{actd}, so that the gravity-induced interaction is negligible and $M$ is actually independent from the Planck mass.

Whatever the interpretation of the four-fermion interaction, and since we have not enough input to \emph{predict} the scale at which condensation takes place, for the time being we must content ourselves to notice that a tiny value of $M$ may be very compelling phenomenologically. In fact, the same gap $\Da\sim M$ may also be able to account for neutrino oscillations which, as is well known, happen at the same mass scale as dark energy~\cite{AK}.

The cosmological evolution of the gap energy is now clear. ``Initially,''  $\Da\sim\mu\gg M$, and the gap energy behaves as negative dark radiation. Provided the very early universe is dominated by an energy density component which redshifts slower than radiation ($w_{\rm eff}<1/3$), such as during inflation ($w_{\rm eff}\approx -1$) or a stringy thermal phase ($w_{\rm eff}\approx 0$)~\cite{h1,h2,h3,h4,h5}, the negative gap energy ensures the existence of a nonsingular bounce point where the gap energy density precisely cancels that of ordinary matter. After the bounce, the gap energy density redshifts away faster than regular matter and remains  subdominant as compared to ordinary matter/radiation. Thus we can have the usual decelerating phase of the standard cosmological model. However, once $\Da\sim\mu\sim M$, we gradually fall into the constant gap regime discussed above, where $\Da\sim M\gg \mu$. Once the matter energy density drops down to $\rho_{\rm m}\sim M^4$, we enter the present dark energy dominated de Sitter phase.


\subsection{Numerical Explorations}

We will now verify numerically that we indeed obtain the late-time cosmology discussed above, and, in particular, undergo a transition from an early decelerating to a late accelerating phase. To this purpose we define
\be
\phi\equiv -\ln\Da^2\,.
\ee
The gap equation becomes
\be\label{gapeq}
2\pi^2M^2=\rme^{-\phi}(N+1-\phi)-2\mu^2(N+2-\phi)\,.
\ee
From Eq.~\Eq{neq} one obtains
\be\label{deneq}
n=\frac{n_0}{a^3}=-\frac{\mu}{2\pi^2} \rme^{-\phi}(N+1-\phi)\,.
\ee
Without loss of generality we fix $n_0=1/(2\pi^2)$. Inverting with respect to $\mu$,
\be\label{mueq}
\mu=-\frac{1}{a^3}\frac{\rme^\phi}{N+1-\phi}\,,
\ee
while the gap density is given by
\be\label{vmineq}
\rho_{\rm gap}=\frac{\rme^{-\phi}}{32\pi^2}\LF\rme^{-\phi}-8\mu^2\RF \LF 2N+3-2\phi\RF\,.
\ee
The chemical potential is negative as long as $\phi>-1$ and $N>0$. Assuming that $\dot\phi>0$ (the gap $\Delta$ decreases in time) $|\mu|$ decreases as well, as we shall see later.

\begin{figure}
\includegraphics[width=8.6cm]{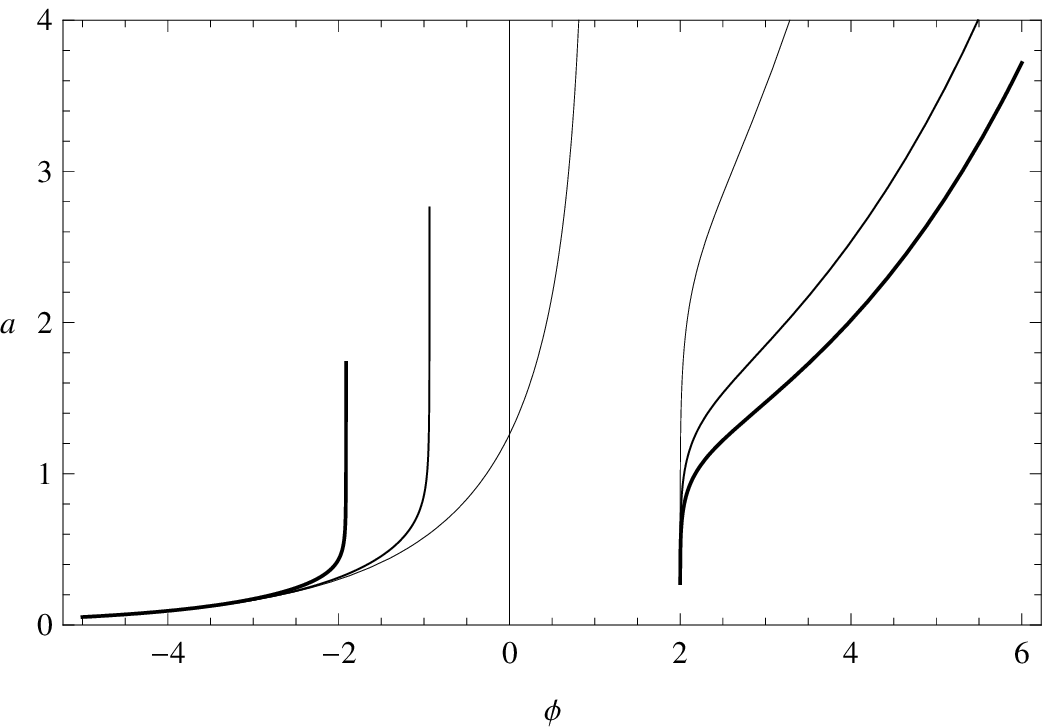}
\includegraphics[width=8.6cm]{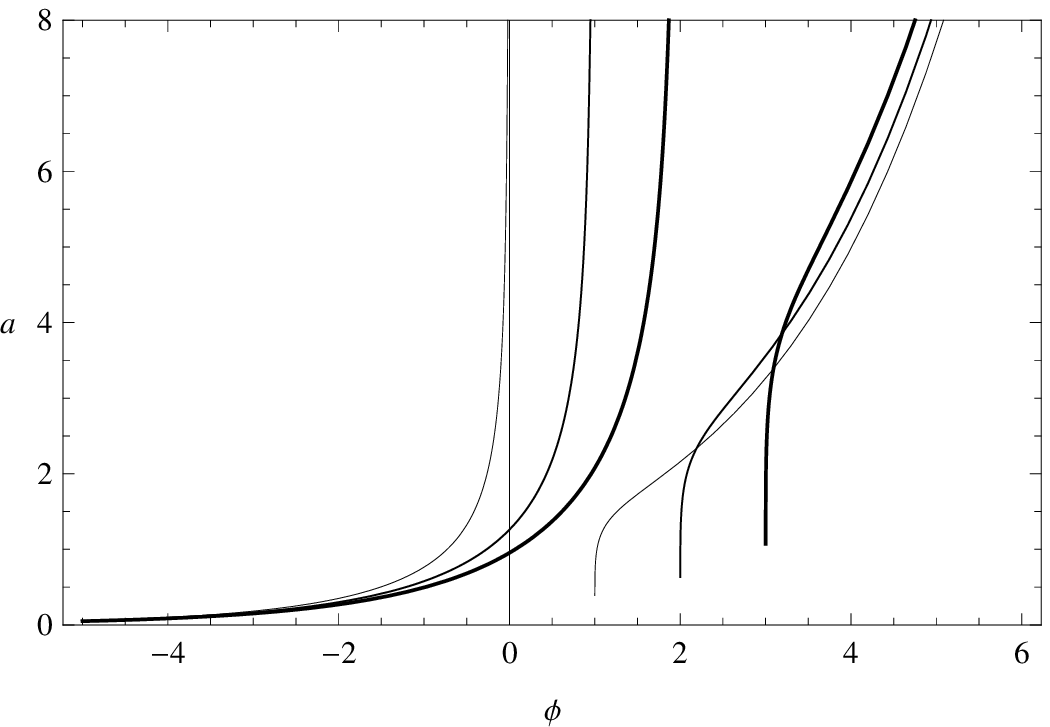}
\caption{\label{fig1} Cosmological branches. Left panel: $N=0$ and $M=0,0.5,1$ (increasing thickness). Right panel: $M=0$ and $N=-1,0,1$ (increasing thickness).}
\end{figure}

Plugging Eq.~\Eq{mueq} in \Eq{gapeq} and solving for the scale factor, one gets
\be
a= \left[\frac{2\rme^{3\phi}(N+2-\phi)}{(N+1-\phi)^2(N+1-\phi-2\pi^2\rme^\phi M^2)}\right]^{1/6},\label{a}
\ee
where we chose the real positive root. By numerically inverting (\ref{a}) we  can identify two disconnected branches for $\phi(a)$ as depicted in Fig.~\ref{fig1}.
\begin{itemize}
\item[(A)] A branch confined within the interval $-\infty<\phi<\phi_{\rm A}$, where $\phi_{\rm A}$ solves the equation $N+1-\phi_{\rm A}-2\pi^2\rme^{\phi_{\rm A}} M^2=0$; \item[(B)] A branch confined within the interval $\phi_{\rm B}\equiv N+2<\phi<+\infty$.
\end{itemize}
From Eq.~(\ref{neq}), we also observe that while in branch (A) $\mu<0$, in the (B) branch $\mu>0$. This suggests that while branch (A) is relevant for strong coupling and BCS condensation, branch (B) describes solutions in the weak-coupling regime where the gap is exponentially suppressed with respect to the chemical potential.

Differentiating Eq.~(\ref{a}) with respect to time, we find
\be
H=\left(2+\frac{N+2-\phi}{N+1-\phi-2\pi^2M^2\rme^\phi}+\frac{2}{N+1-\phi}-\frac{1}{N+2-\phi}\right)\frac{\dot\phi}{6}\,.
\ee
The two branches are separated by a singularity since the Hubble rate diverges as $\phi\ra\phi_{\rm A}$ or $\phi\ra\phi_{\rm B}$. As $\phi_{\rm A}<N+1<\phi_{\rm B}$, we cannot go from one to the other. Thus, identifying the relevant branch is important for the choice of initial conditions, as well as the ensuing cosmology. Figure \ref{fig1} shows how the branches change with respect to $M$ and $N$: $M$ determines $\phi_{\rm A}$ and the increase rate of the scale factor, while $N$ determines only $\phi_{\rm A}$ and $\phi_{\rm B}$. This is reassuring as the parameter $N$ is arbitrary and its role is just a $\phi$ translation, although its actual value does determine the physical scale of $\rho_{\rm gap}$.


Since we are interested in the (A) branch with negative chemical potential representing the BCS condensation phase, we shall evolve the equations of motion from initial conditions typical of this branch. For the extra matter component we consider nonrelativistic dust, for the purpose of illustration. Thus we set $\rho_{\rm m}=a^{-3}$, $M=0$ (strong coupling) and $N=0$. These values will not correspond to the observed universe but they will capture the qualitative features of the dark energy solution. The acceleration of the universe is encoded in the first slow-roll parameter
\be
\epsilon\equiv -\frac{\dot H}{H^2}\,.
\ee
$\epsilon>1$ corresponds to a decelerating universe, while $\epsilon<1$ signals acceleration, if the universe expands.

\begin{figure}
\includegraphics[width=7.4cm]{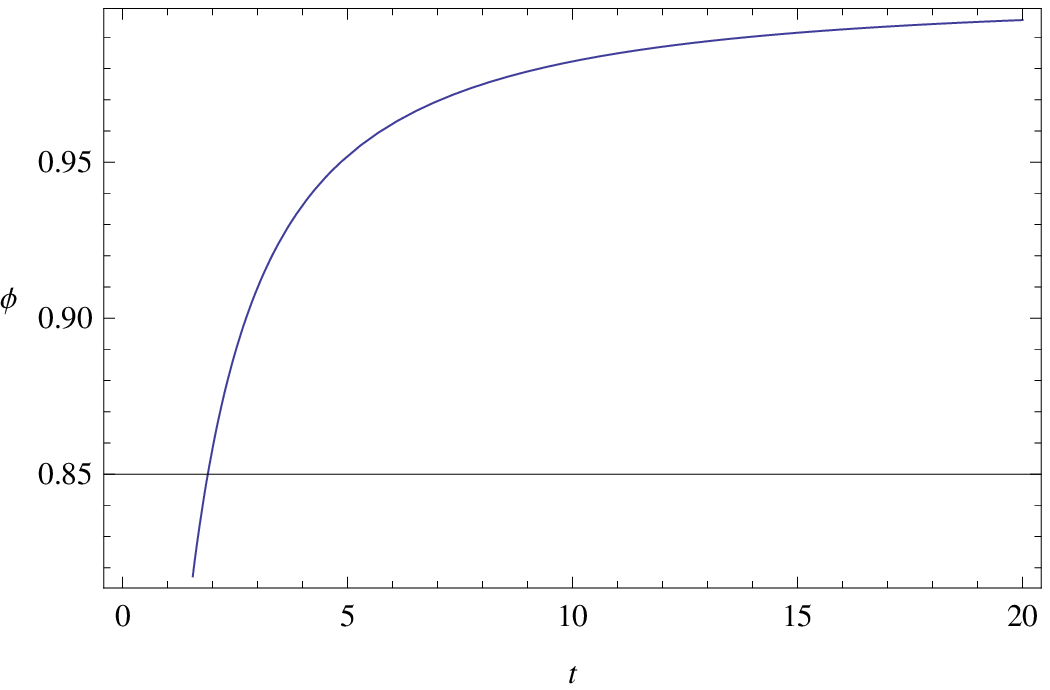}
\includegraphics[width=7.4cm]{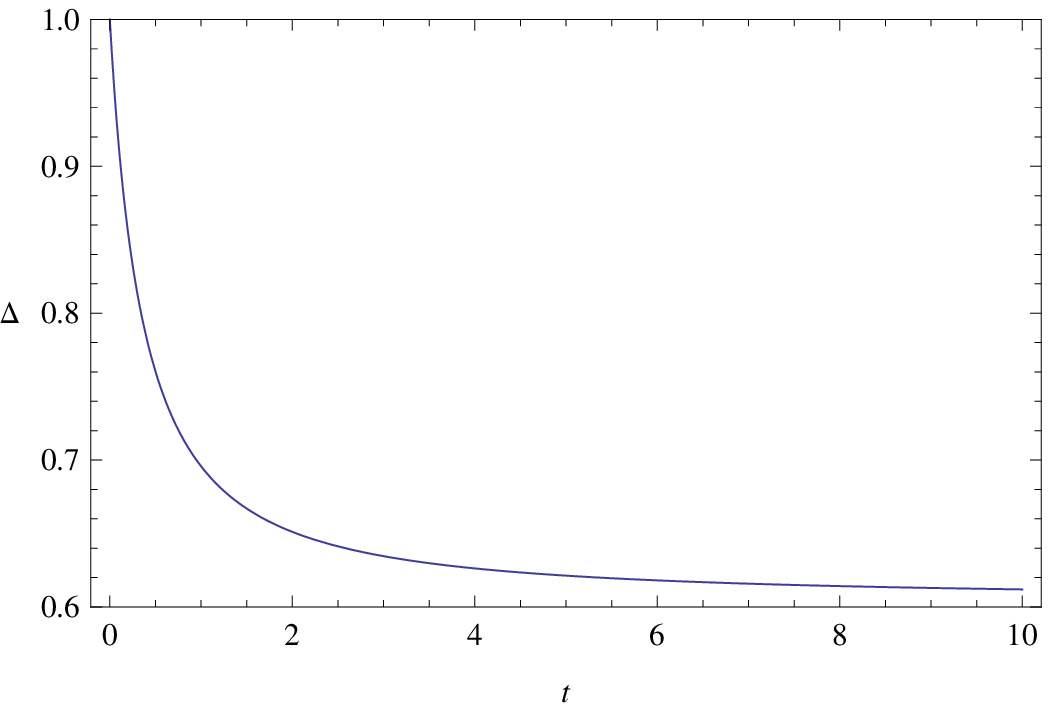}
\includegraphics[width=7.4cm]{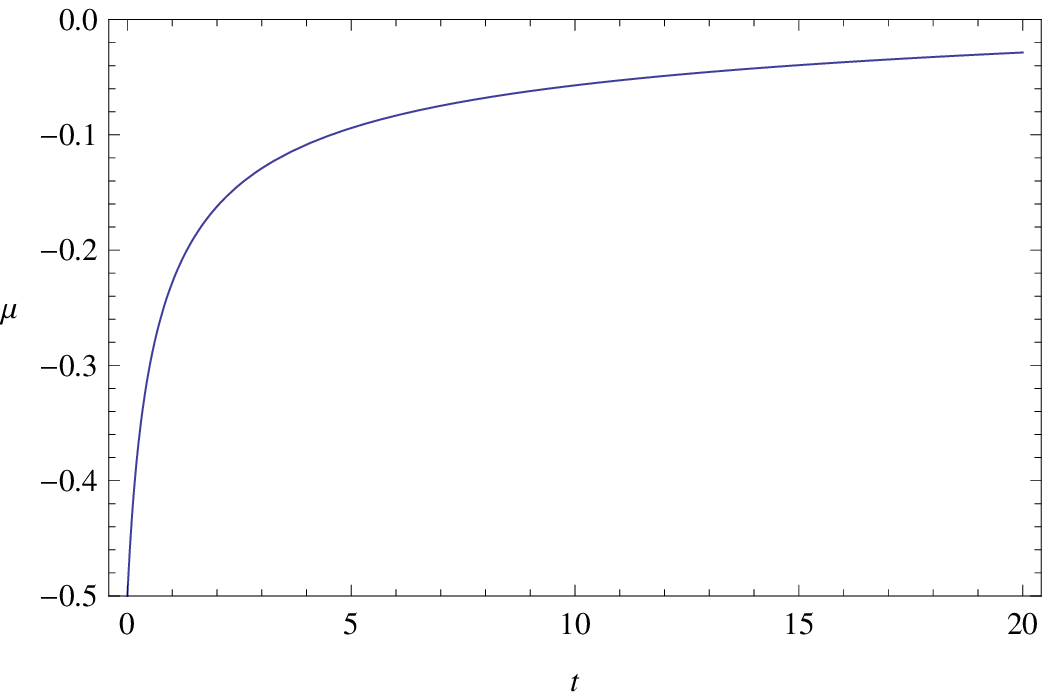}
\includegraphics[width=7.4cm]{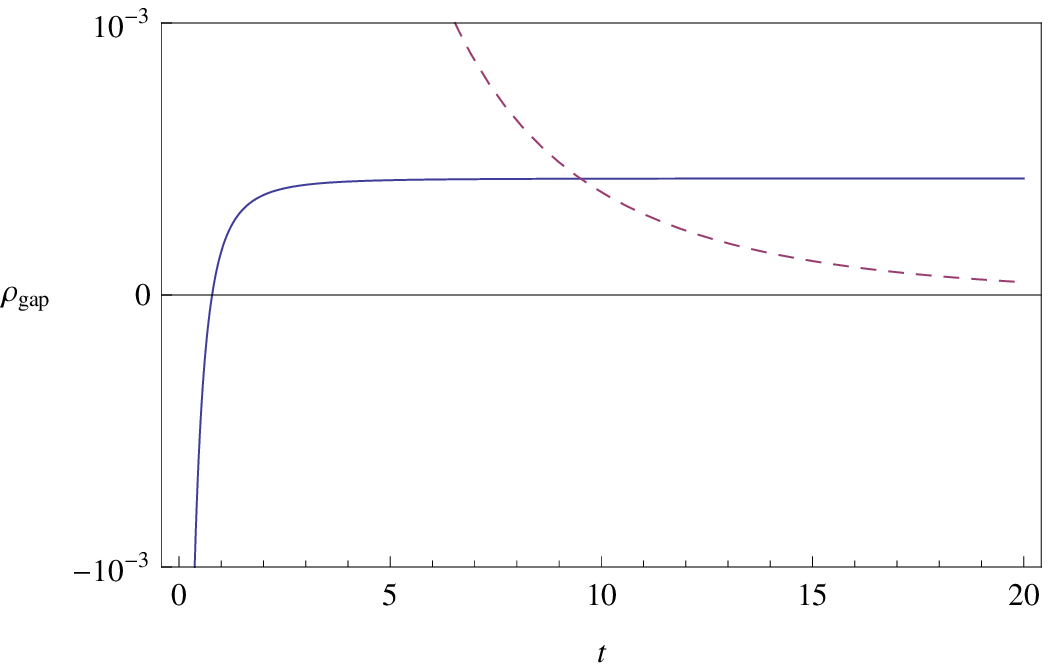}
\caption{\label{fig2}
Numerical late-time solution for $M=0=N$. From left to right and top to bottom: $\phi(t)$, gap $\Delta(t)$, chemical potential $\mu(t)$ and gap energy density $\rho_{\rm gap}(t)$. In the last panel the matter contribution (dashed line) is shown for comparison.}
\end{figure}

In Fig.~\ref{fig2} we show the evolution of the gap and its energy density in synchronous time. As $\phi\to 1$, $\Da\to \rme^{-1/2}\approx 0.6065$. The gap $\Da$ decreases in time, so that the difference between the Fermi sea energy and the true vacuum of the theory becomes negligible. Initially, in the decelerating phase  $|\mu|\sim \Da$ as we expected from analytical arguments. However, at later times, $\mu\to 0^-$, meaning that the density of Cooper pairs (which is positive, consistently) decreases; this is because the formation of pairs becomes less and less favorable. Initially $\rho_{\rm gap}<0$ and the matter contribution dominates. The gap density, however, {\it increases} in time and eventually (after having changed sign) dominates over dust. When this happens we enter the late-time asymptotic de Sitter phase.

\begin{figure}
\includegraphics[width=7.4cm]{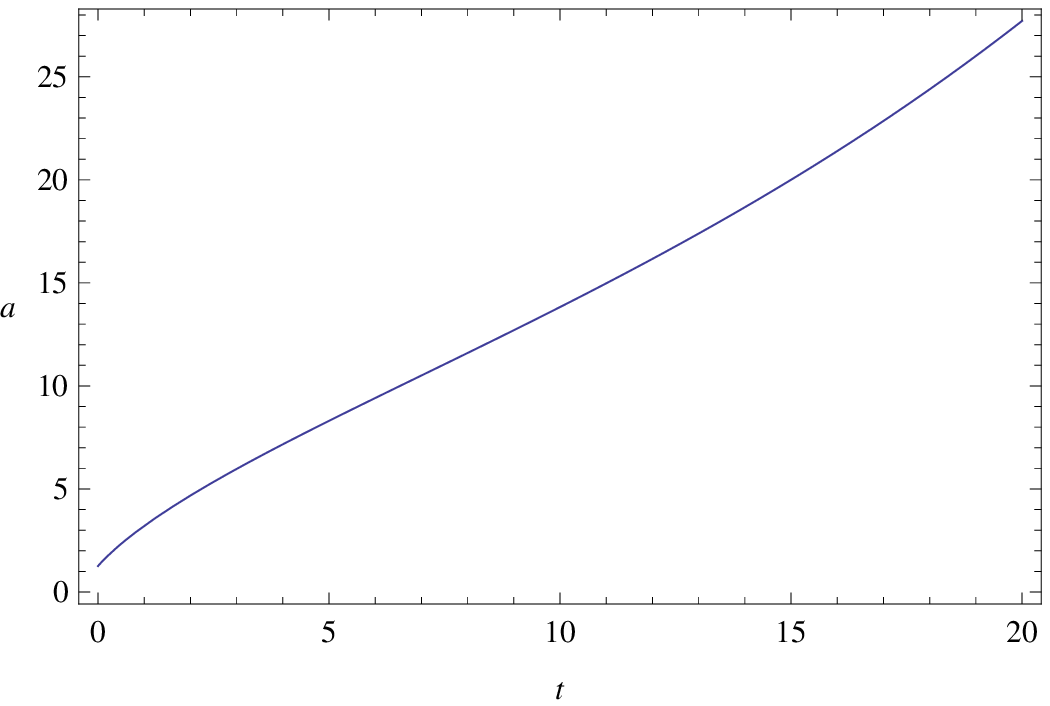}
\includegraphics[width=7.4cm]{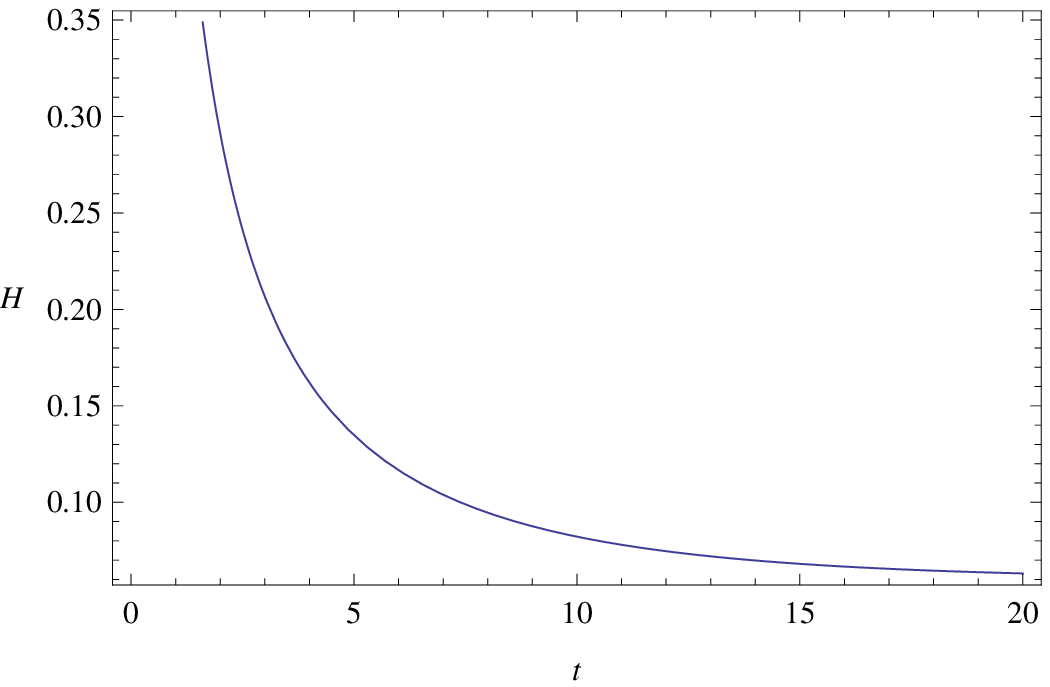}
\includegraphics[width=7.4cm]{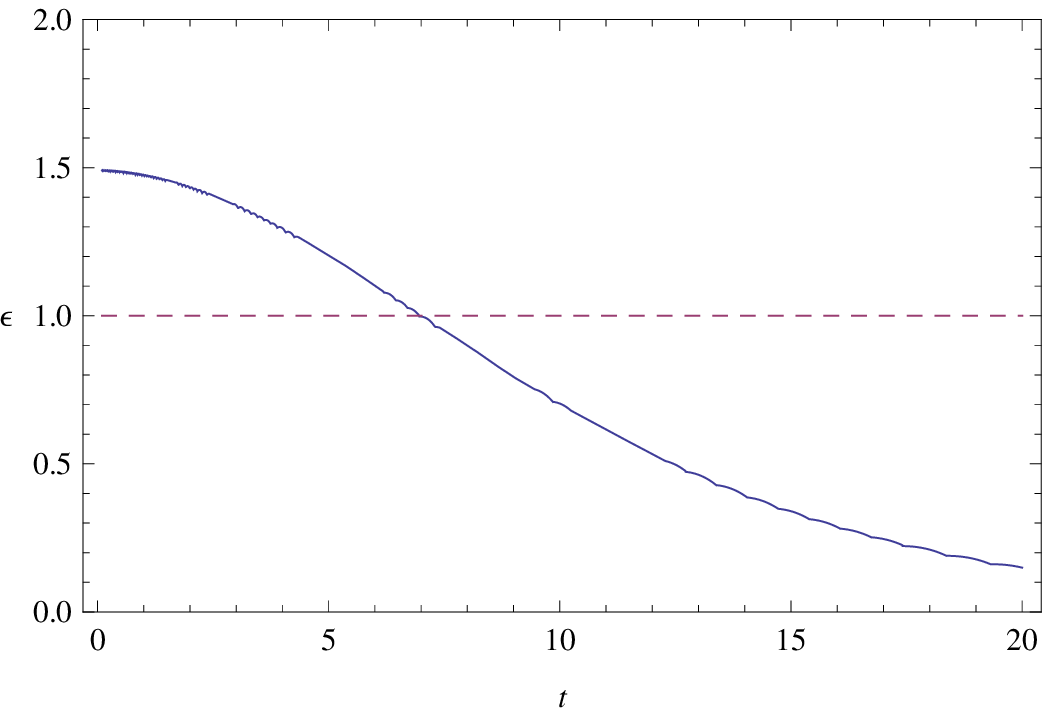}
\includegraphics[width=7.4cm]{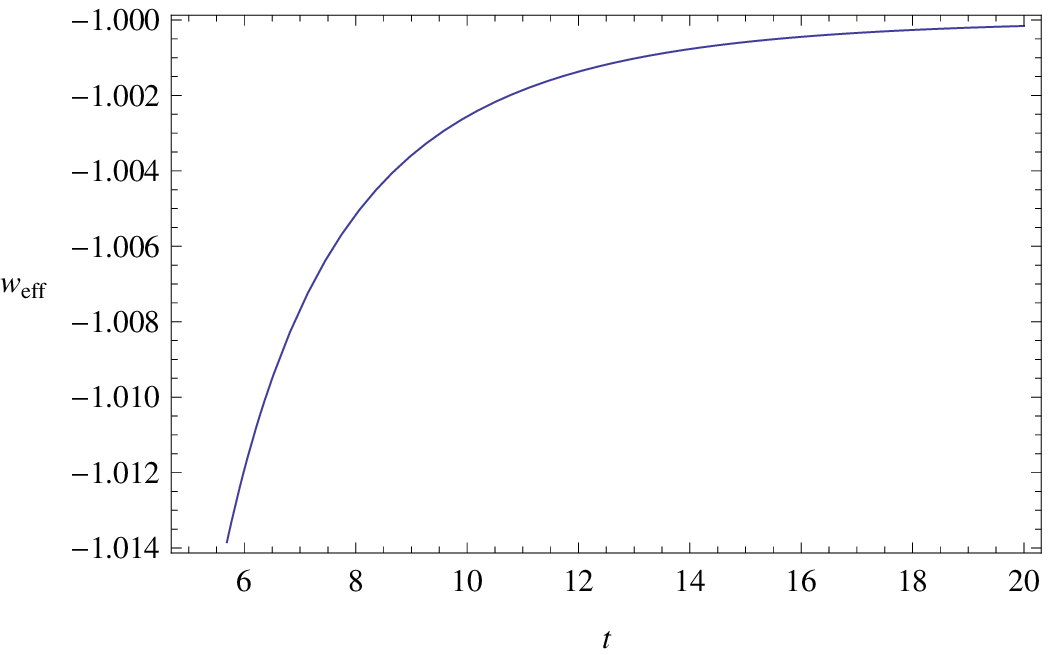}
\caption{\label{fig3} Numerical late-time solution for $M=0=N$. From left to right and top to bottom: $a(t)$, Hubble parameter $H(t)$, first slow-roll parameter $\epsilon(t)$ and effective barotropic index $w_{\rm eff}(t)$ of the condensate equation of state.}
\end{figure}

The evolution of the scale factor and its derivatives is plotted in Fig.~\ref{fig3}. At early times the universe is dominated by the pressureless matter component $\rho_{\rm m}$ and in fact $a\sim t^{1/\epsilon}$, where $\epsilon\sim 3(1+w)/2\sim 3/2$. As one can see from the last plot, acceleration is triggered when $\rho_{\rm gap}\sim\rho_{\rm m}$ but slightly \emph{before} $\rho_{\rm gap}>\rho_{\rm m}$. When radiation is also added, we have checked that the early-time behaviour changes accordingly, $\epsilon\sim 2$, but the overall picture remains the same.

The effective equation of state of the condensate, Eq.~\Eq{weff}, is shown in the last panel of Fig.~\ref{fig3}. The plot starts from $t=\cO(10)$ because, in accordance with Fig.~\ref{fig2}, matter dominates before that time. A characteristic prediction of this solution, which can be tested by future observations, is that $w_{\rm eff}<-1$ and $\dot{w}_{\rm eff}>0$ at late times. The actual value of the barotropic index at the onset of the dark energy era will depend on the couplings of the model, which we have not tuned with respect to observations. Since we have a phantomlike regime at intermediate-to-late times, one should be able to distinguish our model from a pure cosmological constant or quintessence field at high enough redshift. The fit with the supernov\ae\ data goes beyond the goals of the present investigation.


\section{Discussion}\label{disc}

In this paper we have proposed a mechanism realizing late-time cosmic acceleration due to the possibility that fermions can condense in the very early universe. We have shown that all of the necessary conditions for condensation exist, including a covariant attractive four-fermion interaction. By analyzing the transcendental equation relating the gap and the scale factor we discovered that at late times the contribution of the gap energy drives a phase of acceleration. The details of the model depend on the regularization scheme and the value of the couplings, but the emerging qualitative picture is robust.

In this context, we note that one may be worried that, given the low value of $M$ required for phenomenology, $M\sim {\rm meV}$, our procedure for integrating out gravitational degrees of freedom is not justified, and whether we might be missing corrections of order $\Delta/M$, which may be important as $\Delta/M$ is not small. Let us reemphasize why the method we adopted to obtain the effective potential for $\Delta$ is robust. In the first step, we integrated out the torsion field to obtain a four-fermion interaction. This was done by imposing the classical equations of motion for the torsion field. However, we note crucially that the torsion is a nondynamical field: its field equation is algebraic and there are no derivatives acting on the torsion. In other words, the torsion field does not propagate. If one were to draw  Feynman diagrams involving  torsion propagators, these would just collapse into  points leading to four-fermion interaction vertices. Unlike gauge theories where the four-fermion interaction is only an ``effective'' interaction valid below the mass of the gauge bosons, the torsion-mediated four-fermion interaction is an exact quantum mechanical description, because the torsion field has no kinetic term. Therefore the actions given by \Eq{faction} and \Eq{actd} are exactly equivalent to \Eq{interaction} (plus the usual Einstein--Hilbert and Dirac actions) both classically and quantum mechanically.  This equivalence does not depend on how small or large the value of $M$ is.

In the next step, we integrated out the fundamental fermionic degrees of freedom to obtain an effective theory of their bound state, the gap $\Delta$ (Cooper pairs). The procedure we followed is precisely the same that is adopted in condensed matter literature. In particular, this is known to capture not only the weak-coupling BCS limit ($M \ra \infty$), but also the strong-coupling BEC limit (Bose--Einstein Condensate, $M \ra 0$), which is what we are interested in (for general reviews on BEC and the BEC--BCS crossover, see \cite{DGP,leg01,PeS,PiS,GPS}). We note that the effective potential for $\Delta$ features a $\ln (\Delta/M)$ term, which contains all the powers of $(\Delta/M)$. In this sense, the effective action $S_{\rm eff}$ in Eq.~\Eq{effa} describes nonperturbative phenomena by summing all such terms.

Having clarified the nonperturbative character of the condensation, it remains to explain the small value of $M$, which is a partially unrelated question. We have not answered it here but we have commented on several possibilities, including a renormalization mechanism and the presence {\it ab initio} of a strong fermionic interaction. To summarize, one is indeed assuming that a bare $\cO(1)$ cosmological constant ``miraculously cancels'' by virtue of some mechanism. So, we do not solve the ``old cosmological constant problem'' and in this sense fair no better than the usual quintessence models, where one assumes that all contributions to the effective cosmological constant conspire to reduce it to a dynamical field with a ``small'' potential. However, our model has some distinct advantages. First, because the suppression of the cosmological constant is exponential, the level of fine-tuning is reduced to just one part over 100, via the choice of $N$. Therefore, if one could give a physical interpretation to $N$, then  the ``smallness problem'' associated with the observed value of dark energy would be relaxed. This was attempted in Sec.~\ref{epl}. Second, our situation is consistent because we have calculated the nonperturbative potential, whereas in usual quintessence models it is not clear why higher-order corrections are suppressed. 

There are several issues we have not considered. Since we have worked on the assumption that, like in Minkowski, the fermion interaction is nonrenormalizable, the regularization parameter $N$ of the effective theory has been assumed to be physical and therefore should ultimately be motivated by the fundamental microscopic theory. Also, we have assumed $N$ to be a constant. A possibility we leave for future study is to allow for a time-varying $N(t)$ or coupling $M(t)$. For instance, previous literature~\cite{CR,GHN,MW} have considered different choices for a physical cutoff,  $\la=\mu$ or $\la=H$, which would correspond to having a time-varying $N$.\footnote{The ambiguity in the choice of the cutoff is not dissimilar to the one entailed in modern and inequivalent formulations of loop quantum cosmology \cite{Ash07,Boj08}. There, one can choose the kinematical area of the  elementary holonomy to be fixed in time and equal to the Planck area (improved quantization scheme \cite{APS}), or else make it dynamical as in lattice refinement models \cite{Boj06,BCK,CH}, which corresponds to probing geometry with a time-dependent microscope.} This may lead to interesting situations, including the possibility of not having to tune $M$ to the very tiny meV scale.

In order to verify the robustness and observational validity of the rich cosmological picture we presented here, future studies will have to go into greater detail in the analysis of the parameter space and initial conditions. It is promising that the same condensation mechanism can solve the big bang singularity and the dark energy problem. We end by pointing out that the dark energy scenario presented above has some distinctive observational features. For instance, depending upon the detailed history of the early universe the negative dark radiationlike gap energy may be detectable in BBN and CMB observations~\cite{radiation1,radiation2}. Also, there is a most encouraging possibility of linking the scenario with neutrino physics~\cite{AK}, as mentioned earlier.


\begin{acknowledgments}
S.A. and T.B. are supported by an NSF CAREER grant. G.C. is supported by NSF Grant No.~PHY0854743, the George A. and Margaret M. Downsbrough Endowment and the Eberly research funds of Penn State. T.B. would like to acknowledge the hospitality of the physics department at University of Minnesota in Minneapolis.
\end{acknowledgments}


\appendix

\section{Derivation of the four-fermion interaction}

The Holst action can be written as ($G=1$)
\be
S_{\rm H} = \frac{1}{16\pi}\int\rmd^{4}x\,e\,e^{\mu}_{I}e^{\nu}_{J}P^{IJ}{}_{KL}R^{\ \ KL}_{\mu\nu}\,,
\ee
where
\be
P^{IJ}{}_{KL} = \delta^I_{[K}\delta^J_{L]} - \frac{1}{2\gamma}\epsilon^{IJ}{}_{KL},
\ee
whose inverse is
\be
P^{-1}{}_{IJ}{}^{KL} = \frac{\gamma^2}{\gamma^2+1}\left( \delta^K_{[I}\delta^L_{J]}+ \frac{1}{2\gamma}\epsilon_{IJ}{}^{KL}\right).
\ee
Variation of the Holst action with respect to the connection yields
\be
\frac{\delta S_{\rm H}}{\delta A_\nu^{\ KL}} = -\frac{1}{8\pi}D_\mu \left(
    e\,e^{[\mu}_{I}e^{\nu]}_{J}\right)P^{IJ}{}_{KL}\,.
\ee
Likewise, variation of the Dirac action gives
\ba
    \frac{\delta S_D}{\delta A_\nu^{\ KL}} &=& - \frac{\rmi}{8}e\bar\psi\{\gamma_{[K}\gamma_{L]},\gamma^I\}e^\nu_I\psi \nonumber\\
            &=& \frac{e}{4}\epsilon^I{}_{KLM}(\bar\psi\gamma_5\gamma^M\psi)e^\nu_I.
\ea
In the second line we have used the identity $\{\gamma_{[K}\gamma_{L]},\gamma_I\} =
2\rmi\epsilon_{IKLM}\gamma_5\gamma^M$. The total variation of the action $S_{\rm H}+S_{\rm D}$ with respect to the connection is
\be\label{Gauss_Law}
D_\mu \left(e\,e^{[\mu}_{I}e^{\nu]}_{J}\right)P^{IJ}{}_{KL} = 2\pi e\epsilon^I{}_{KLM}J_5^M e^\nu_I\,,
\ee
where $J_5^M=\bar\psi\gamma_5\gamma^M\psi$ is the axial current.

Writing the connection as $A_\mu^{IJ} = \omega_\mu^{IJ} + C_\mu^{IJ}$, where $\omega$ is the
connection compatible with the tetrad, and using Eq.~(\ref{Gauss_Law}), one gets
\be
C_{\mu[P}{}^\mu e^\nu_{Q]} + C_{[PQ]}{}^\nu = 2\pi\frac{\gamma^2}{\gamma^2+1}e^\nu_I J_{5\,M}\left(\epsilon^{MI}{}_{PQ} + \frac{1}{\gamma}\delta^M_{[P}\delta^I_{Q]} \right)\,.
\ee

Contracting with $e_\nu^P$ we obtain
\be\label{vecax}
C_{\mu Q}{}^\mu = \frac{3\pi}{8}\frac{\gamma}{\gamma^2+1}J_{5\,Q}\,.
\ee
From the above two equations we obtain
\be
C_\mu^{\ IJ} = 2\pi\frac{\gamma^2}{\gamma^2 + 1}J^M_5\left(\epsilon_{MK}{}^{IJ}e^K_\mu
               -\frac{2}{\gamma}\delta^{[J}_M e^{I]}_\mu\right).
\ee
Inserting the above expression into the first-order gravity + matter action yields the four-fermion interaction. For the calculation in the Nieh--Yan case we refer the reader to \cite{Mer06,MT}.


\end{document}